\documentstyle[12pt,aaspp4]{article}
\begin{document}
\newcommand{\up}[1]{\ifmmode^{\rm #1}\else$^{\rm #1}$\fi}
\newcommand{\zdot}{\makebox[0pt][l]{.}}
\newcommand{\upd}{\up{d}}
\newcommand{\uph}{\up{h}}
\newcommand{\upm}{\up{m}}
\newcommand{\ups}{\up{s}}
\newcommand{\arcd}{\ifmmode^{\circ}\else$^{\circ}$\fi}
\newcommand{\arcm}{\ifmmode{'}\else$'$\fi}
\newcommand{\arcs}{\ifmmode{''}\else$''$\fi}

\title{The Araucaria Project. OGLE-LMC-CEP-1718: An exotic eclipsing binary system
composed of two classical overtone Cepheids in a 413-day orbit
\footnote{Based in part on observations obtained with the ESO 3.6m telescope for 
Programme 091.D-0469(A). This paper includes data gathered with the 6.5m
Magellan Clay Telescope at Las Campanas Observatory, Chile.}
}
\author{Wolfgang Gieren}
\affil{Universidad de Concepci{\'o}n, Departamento de Astronomia,
Casilla 160-C, Concepci{\'o}n, Chile}
\authoremail{wgieren@astro-udec.cl}

\author{Bogumi{\l} Pilecki} 
\affil{Universidad de Concepci{\'o}n, Departamento de Astronomia,
Casilla 160-C, Concepci{\'o}n, Chile} 
\affil{Warsaw University Observatory, Al. Ujazdowskie 4, 00-478, Warsaw,
Poland}
\authoremail{bpilecki@astro-udec.cl}

\author{Grzegorz Pietrzy{\'n}ski}
\affil{Universidad de Concepci{\'o}n, Departamento de Astronomia,
Casilla 160-C, Concepci{\'o}n, Chile}
\affil{Warsaw University Observatory, Al. Ujazdowskie 4, 00-478, Warsaw,
Poland}
\authoremail{pietrzyn@astrouw.edu.pl}

\author{Dariusz Graczyk}   
\affil{Universidad de Concepci{\'o}n, Departamento de Astronomia,
Casilla 160-C, Concepci{\'o}n, Chile}
\authoremail{darek@astro-udec.cl}

\author{Ian B. Thompson}
\affil{Carnegie Observatories, 813 Santa Barbara Street, Pasadena, CA 91101-1292, USA}
\authoremail{ian@obs.carnegiescience.edu}

\author{Igor Soszy{\'n}ski}
\affil{Warsaw University Observatory, Al. Ujazdowskie 4, 00-478, Warsaw, Poland}
\authoremail{soszynsk@astrouw.edu.pl}

\author{Piotr Konorski}   
\affil{Warsaw University Observatory, Al. Ujazdowskie 4, 00-478, Warsaw, Poland}
\authoremail{piokon@astrouw.edu.pl}

\author{Rados{\l}aw Smolec}   
\affil{Copernicus Astronomical Centre, Polish Academy of Sciences, Bartycka 18, 00-716 Warsaw, Poland}
\authoremail{smolec@camk.edu.pl}

\author{Andrzej Udalski}
\affil{Warsaw University Observatory, Al. Ujazdowskie 4, 00-478, Warsaw, Poland}
\authoremail{udalski@astrouw.edu.pl}

\author{Nicolas Nardetto}
\affil{Laboratoire Lagrange, UMR7293, Universit{\'e} de Nice Sophia-Antipolis, CNRS, Observatoire de la
Cote d'Azur, Nice, France }
\authoremail{Nicolas.Nardetto@oca.eu}

\author{Giuseppe Bono}
\affil{Dipartimento di Fisica, Universita di Roma Tor Vergata, Via della Ricerca Scientifica 1, 00133 Roma, Italy}
\affil{INAF-Osservatorio Astronomico di Roma, Via Frascati 33, 00040 Monte Porzio Catone, Italy}
\authoremail{Giuseppe.Bono@roma2.infn.it}

\author{Pier Giorgio Prada Moroni}
\affil{Dipartimento di Fisica, Universita di Pisa, Largo B. Pontecorvo 3, 56127 Pisa, Italy}
\affil{INFN, Sezione di Pisa, Largo B. Pontecorvo 3, I-56127, Italy}
\authoremail{prada@df.unipi.it}

\author{Jesper Storm}
\affil{Leibniz-Institut fuer Astrophysik Potsdam (AIP), An der Sternwarte 16, D-14482
Potsdam, Germany}
\authoremail{jstorm@aip.de}

\author{Alexandre Gallenne}
\affil{Universidad de Concepci{\'o}n, Departamento de Astronomia, Casilla 160-C,
Concepci{\'o}n, Chile}
\authoremail{dgallenne@astro-udec.cl}

\begin{abstract}
We have obtained extensive high-quality spectroscopic observations of the OGLE-LMC-CEP-1718
eclipsing binary system in the Large Magellanic Cloud which Soszynski et al. (2008) had identified
as a candidate system for containing two classical Cepheids in orbit. Our spectroscopic data clearly
demonstrate binary motion of the Cepheids in a 413-day eccentric orbit, rendering this eclipsing
binary system the first ever known to consist of two classical Cepheid variables. After disentangling
the four different radial velocity variations in the system we present the orbital solution and
the individual pulsational radial velocity curves of the Cepheids. We show that both Cepheids
are extremely likely to be first overtone pulsators and determine their respective dynamical masses, 
which turn out to be equal to within 1.5 \%.
Since the secondary eclipse is not observed in the orbital light curve we cannot derive
the individual radii of the Cepheids, but the sum of their radii derived from the photometry
is consistent with overtone pulsation for both variables. 

The existence of two equal-mass Cepheids in a binary system having different pulsation periods
(1.96 and 2.48 days, respectively) may pose an interesting challenge to stellar evolution and pulsation
theories, and a more detailed study of this system using additional datasets should yield deeper
insight about the physics of stellar evolution of Cepheid variables. Future analysis
of the system using additional near-infrared photometry might also lead to a better
understanding of the systematic uncertainties in current Baade-Wesselink techniques of distance determinations to Cepheid variables.
\end{abstract}


\section{Introduction} 

Classical Cepheids are distance indicators par excellence and a fundamental rung
on the cosmic distance ladder, connecting our Milky Way galaxy to galaxies in the Local Group
and beyond (Freedman et al. 2001; Gieren et al. 2005a, 2006; Pietrzynski et al. 2006; Riess et al. 2011). 
In order to render Cepheids even more robust and reliable distance indicators,
it is imperative to understand their physical and evolutionary properties with the highest possible 
accuracy. In that context, it has been a breakthrough to find classical Cepheids in detached,
double-lined
eclipsing binary systems which permit a determinion of their basic physical parameters much more
accurately than what is possible for any single Cepheid star. In particular, the analysis of the
OGLE-LMC-CEP-0227 system located in the Large Magellanic Cloud (LMC), containing a 
classical Cepheid pulsating with a period of 3.8 days together with a stable red giant in a 310-day 
orbit, has yielded for the first time
a Cepheid mass and radius determination accurate to 1 \%, and valuable independent insight on the 
p-factor
needed for Baade-Wesselink-type analyses (Pietrzynski et al. 2010; Pilecki et al. 2013).
A second eclipsing binary system in the LMC containing an even shorter-period 
classical Cepheid, OGLE-LMC-CEP-1812, was analyzed by Pietrzynski et al. (2011) and again yielded
a very accurate measurement of the dynamical mass of the Cepheid. These two Cepheid mass determinations have gone
a long way to solve the famous Cepheid mass discrepancy problem, leading to improved predictions
of Cepheid masses from stellar pulsation and evolution theories. (Marconi et al. 2013; 
Prada Moroni et al. 2012).

In the present paper, we report on the confirmation of an even more exotic, and so far unique,
eclipsing binary system in the LMC consisting of a {\it pair of classical Cepheids in a 413-day orbit}.
The system, herein named OGLE-LMC-CEP-1718, was discovered and identified as a {\it double Cepheid}
by Alcock et al. (1995). Later Soszynski et al. (2008) found that it also exhibits eclipsing variability, but it was not yet clear if the two Cepheids were indeed gravitationally bound. Our spectroscopic observations of this 
double-lined system over the past years
clearly show that the two Cepheids orbit each other, with the additional radial velocity
variability of the Cepheids due to their pulsations  superimposed on their orbital radial velocity curves.
Evidently, the analysis of this system
and the characterization of the physical properties of its coeval Cepheids holds great promise
to deepen our understanding of Cepheid physics and evolution.

In this paper, we present spectroscopic observations of OGLE-LMC-CEP-1718 and extract the orbital radial velocity curves of the two components of the system and the individual pulsational radial velocity curves of the Cepheids. 
We add new photometric data from OGLE III and OGLE IV surveys to that presented by Soszynski et al. (2008) to
the radial velocity data to obtain the orbital solution as well as a determination of several physical parameters
of the Cepheids, particularly their masses and pulsation modes.

\section{Observations and Analysis}
Using the MIKE spectrograph at the 6.5-m Clay Telescope at Las Campanas Observatory and the HARPS
spectrograph attached to the 3.6-m telescope at the ESO La Silla Observatory we obtained 38
high-resolution spectra of OGLE-LMC-CEP-1718 (mean magnitudes I=14.511, V=15.190; Soszynski
et al. 2008) between  September 29, 2011 and December 12, 2013. Radial velocity
determinations from these spectra were made using the Broadening Function Method (Rucinski 1992,
1999) implemented in the RaveSpan code (Pilecki et al. 2012). 
We analyzed the spectra in the range of 4125 to 6800 $\mathring{A}$ using 
as templates the theoretical spectra taken from the library of Coelho et al. (2005). 
The radial velocities determined in this way were typically accurate to 0.3 km/s and 
we have never seen any systematic difference between data obtained with MIKE and HARPS spectrographs.
The measured radial velocities of both components are presented in Table 1.

In addition to these radial velocity data, we were able to use 1535 I-band measurements of the
system collected by the OGLE Project (Udalski 2003) over many years, including data from the
most recent OGLE III and IV surveys. The
previous photometric analysis performed by Soszynski et al. (2008) had detected three periods-
one describing the orbital motion of the two stars, and two additional magnitude variations due to
the pulsations of two Cepheids with periods of 1.96 and 2.48 days. This was a promising indication
that the Cepheids might be gravitationally bound, but it could also be a blend. 

The analysis of the radial velocity dataset confirms the genuine spectroscopic
binary nature of the system, with orbital motion of the two stars superimposed on the intrinsic
radial velocity variations due to their pulsations. Knowing the pulsation periods from the photometry
we were able to fit a complex model of orbital motion together with a pulsation variability for each
object (see also Pilecki et al. 2013). This analysis clearly shows that the orbital period
of the system is only half of the one originally assumed by Soszynski et al. (2008). This is a
consequence of the fact
that with the inclination and orbit orientation of the system, presented in Table 2 together
with other orbital parameters derived in our orbital solution,
only the primary
eclipse is visible in the photometry whereas a possible secondary eclipse remains elusive 
and cannot be detected in the data. The absence of a secondary eclipse 
is fully consistent with the eccentricity and the other orbital parameters obtained in our analysis.

We were able to determine the four individual radial velocity curves 
which are superimposed in the data - the orbital radial velocities of the components of the system, 
and the pulsational 
radial velocity curves of the primary (P=1.96-day) and  secondary (P=2.48-day) Cepheid components
in the system. The orbital radial velocity curve of OGLE-LMC-CEP-1718, and the pulsational radial
velocity curves of its two Cepheids are shown in Figures 1-3. The disentangling of the four different
radial velocity curves is not yet perfect, but the small residuals from the fitted curves which
are only slightly larger than the typical precision of the individual radial measurements demonstrate that the disentangling has been achieved with a high degree of accuracy with our code.

The pulsational radial velocity curves are both low-amplitude and approximately sinusoidal, as are the corresponding I-band light curves of the Cepheids which are shown in Figures 4 and 5.
In Fig. 6 we show the orbital I-band light curve based on the full dataset,  folded on the  orbital period of 412.807 days.

\section{Results and Discussion}
From our orbital solution, we find that the masses of the Cepheids are 3.3 and 3.28 $M_\odot$, respectively, individually determined with an accuracy of 3 \% (see Table 3). Because the high eccentricity error does not contribute to the evaluation of the mass ratio, it is determined with a much better accuracy. The analysis indicates that the two Cepheids in the OGLE-LMC-CEP-1718 system have equal masses to within $\pm$ 1.5 percent.

The very short pulsation periods suggest pulsation in non-fundamental modes. 
In Figure 7 we have plotted the positions of the two Cepheids on the
I-band light curve Fourier decomposition diagrams of Soszynski et al. (2008). The loci of both stars
on these diagrams, particularly on the $R_{21}-\log P$ diagram, strongly suggest that both Cepheids in 
OGLE-LMC-CEP-1718 are pulsating in the first overtone mode. 

This result can be checked in a different way. Since we cannot observe the secondary eclipse
in the light curve, we cannot determine the individual radii of the Cepheids. However the analysis of the light curve does return the sum of the radii, in this case 52.5 $\pm$ 1.5 $R_{\odot}$. Assuming first overtone pulsation for the two Cepheids in OGLE-LMC-CEP-1718, we can calculate their expected radii from a period-radius relation calibrated
for first overtone Cepheids.
Using the observational relation given by Sachkov (2002), we obtain radii of R = (26.1 $\pm$ 2.5) $R_{\odot}$ for the primary, and R = (31.0 $\pm$ 2.6) $R_{\odot}$ for the secondary (longer-period) Cepheid, with the radii ratio of $1.19$.
These predictions are in excellent agreement with the predictions from the theoretical period-radius relation
for first overtone Galactic Cepheids of Bono et al. (2001; 27 and 32 $R_{\odot}$, respectively).
As the ratio is much better constrained than the radii themselves we have used it to calculate
the individual radii of the Cepheids (using the known sum), obtaining $R_1=24 R_{\odot}$ and $R_2=28.5 R_{\odot}$ which is clearly consistent with the values from the given relation.
If we assume fundamental mode pulsation for both Cepheids and use the fundamental mode Cepheid period-radius relation from Sachkov (2002) (which is very similar to other calibrations of that relation, e.g. Gieren
et al. 1998), the expected sum of the radii is 44.0 $\pm$ 0.3 $R_{\odot}$, indicating that fundamental mode pulsation is much more unlikely.
We have to note however, that in this case the calculations are based on the extrapolation as the periods of our stars are shorter than the shortest one among the stars used to obtain the relation.

Yet another argument supporting the first overtone pulsation hypothesis comes from the observed brightness
of OGLE-LMC-CEP-1718. Using the fitted PL relations for first overtone Cepheids in the LMC from the
OGLE project (Soszynski et al. 2008), the expected apparent magnitudes for the primary Cepheid are
15.439 and 16.110 in I and V bands, respectively, whereas for the secondary Cepheid the corresponding
values are 15.104 and 15.786. This leads to expected total apparent magnitudes of both components of
$I_{tot}$ = 14.506 and $V_{tot}$ = 15.183, respectively. The observed apparent magnitudes of the system are
$I_{obs}$ = 14.511 and $V_{obs}$ = 15.190,  in excellent agreement with the expected magnitudes if both
Cepheids pulsate in the first overtone mode.

One possibility to reconcile the fact that both Cepheids in the system have the same masses, but different pulsation periods,
would be the assumption that the primary, shorter-period Cepheid is actually pulsating in the second overtone mode.
The observed period ratio of 1.96/2.48 = 0.79 would be consistent with the hypothesis that the primary Cepheid
is pulsating in the second overtone while the secondary Cepheid is a first-overtone pulsator-a period ratio of
about 0.8 is indeed commonly observed for double-mode 1O/2O Cepheids. However, in the OGLE database of Cepheids
in the Magellanic Clouds (Soszynski et al. 2008) which contains the largest samples of Cepheids, among others
about 100 single-mode second-overtone Cepheids and about 420 double-mode 1O/2O Cepheids, the largest known period
of a second overtone Cepheid is 1.32 days (in the double-mode Cepheid OGLE-SMC-CEP-0305 in the Small Magellanic Cloud).
This is very much shorter than the period of 1.96 days observed for the primary Cepheid in our OGLE-LMC-CEP-1718
system. 

The largest known amplitude of the I-band light curve of a second overtone Cepheid is 0.138 mag (in the single-mode
Cepheid OGLE-SMC-CEP-3509). Our object has a smaller amplitude of 0.097 mag (see Fig. 4), but the amplitude is
decreased by the light from the secondary component through blending. Transforming the magnitudes to fluxes and removing the
contribution of the secondary Cepheid, and transforming the flux back to magnitudes now yields an I-band amplitude of
0.231 mag for the short-period primary Cepheid in the OGLE-LMC-CEP-1718 system. This is much larger than any known
amplitude of a second-overtone oscillation.

The shapes of the light curves cannot be directly compared because there are no second-overtone Cepheids with periods
around 2 days. However, it can be stated that all second-overtone Cepheid light curves are more symmetrical than 
the one of the primary in our system, they are indeed nearly sinusoidal. Finally, the total luminosity of the
OGLE-LMC-CEP-1718 system perfectly agrees with the assumption that the system consists of two first-overtone Cepheids,
as already mentioned above. In conclusion, if the 1.96-day Cepheid in our system would indeed be a second overtone
pulsator, it would be the longest-period and the largest-amplitude second-overtone Cepheid known in any galaxy. Given
the large number of second-overtone Cepheids known to-date, it seems extremely unlikely that we found such an extreme
object in the only eclipsing binary system consisting of two classical Cepheids that has been discovered so far.

Our conclusion then is that we have found a system composed of two classical Cepheids which have 
within a 1.5 \% uncertainty identical masses, both stars are almost certainly
pulsating in the first overtone mode, presumably have the same ages, but have substantially different periods and
luminosities. It will be challenging for stellar evolutionary theory to explain the observed properties
of these Cepheids, which will be the topic of a forthcoming study of our group.

\section{Summary}

We have presented the first confirmed eclipsing binary system with both components identified
as classical Cepheids. These two variables orbit each other in an eccentric orbit with a period of
413 days. Strong evidence is presented that both variables are pulsating in the first overtone mode.
The dynamical masses of both Cepheids are identical to within 2 percent and the absolute masses are
determined with an accuracy of 3 percent.
Both orbital light and radial velocity curves of the system and the pulsational
light and radial velocity curves of the two Cepheids are very well determined from our data although
improvement is possible and desirable.

The OGLE-LMC-CEP-1718 system, as the first confirmed eclipsing binary system containing two classical Cepheids, might
well turn out to be a Rosetta stone for our deeper understanding of the pulsational and evolutionary
properties of Cepheids. In response to this opportunity, and the challenges presented
by the system we plan to improve the existing radial velocity data, and obtain   
observations of near- and mid-IR photometry which should help to better understand the physics of these variables.
Near-infrared light curves for the two Cepheids in the system will in 
particular  provide a unique opportunity
to test the basic assumptions made in the near-infrared surface brightness technique (Fouqu\'e \& Gieren 1997;
Gieren et al. 2005b; Storm et al. 2011) of the distance determination to Cepheids by taking advantage
of the knowledge that both Cepheids are at the same distance.

\acknowledgments
We gratefully acknowledge financial support for this work from the BASAL Centro
de Astrofisica y Tecnologias Afines (CATA) PFB-06/2007, and from the Millenium Institute of Astrophysics (MAS)
of the Iniciativa Cientifica Milenio del Ministerio de Economia, Fomento y Turismo de Chile, project IC120009.
Support from the Polish National Science Center grant MAESTRO 2012/06/A/ST9/00269
and the TEAM subsidy of the Foundation for Polish Science (FNP) is also acknowledged.
The OGLE project has received funding from the European Research Council
under the European Community's Seventh Framework Programme (FP7/2007-2013) / ERC grant agreement no. 246678 to AU.
AG acknowledges support from FONDECYT grant No. 3130361.
We greatly appreciate the expert support of the staffs at the Las Campanas and ESO La Silla Observatories
where the data for this project were obtained. We thank the referee, Laszlo Szabados, for constructive remarks
which helped to improve this paper.

\begin{figure}[p]
\includegraphics{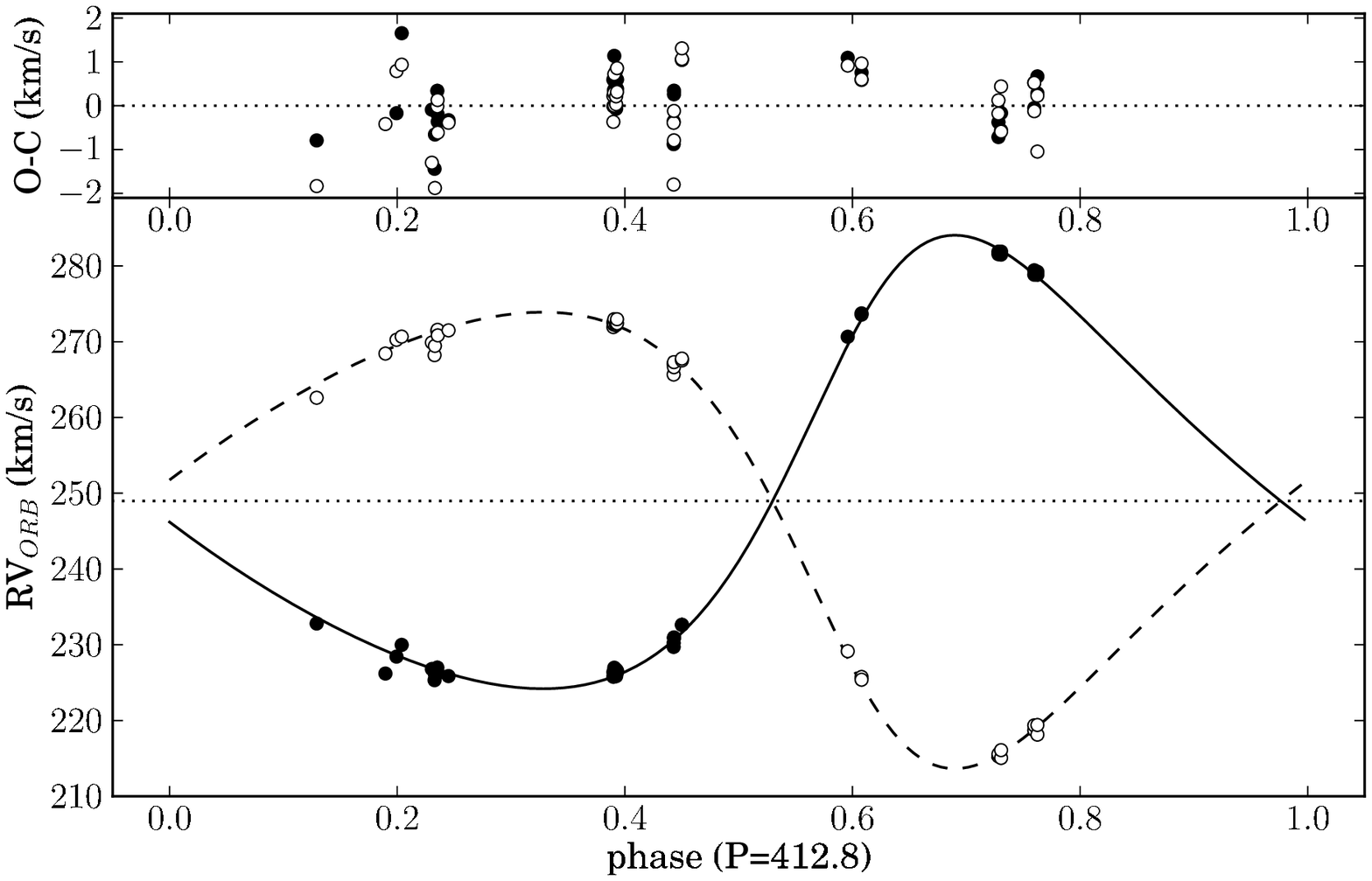}
\vspace{10cm} 
\caption{
 Orbital radial velocity curve of the OGLE-LMC-CEP-1718 system. Filled circles denote the
 primary component (the P=1.96 day Cepheid), open circles the secondary component (the
 P=2.48 day Cepheid). The pulsations of both Cepheids were removed from the observed 
 radial velocities, yielding the pure orbital motion of the stars. The plot is based on the 38
 individual radial velocity observations of the system reported in Table 1.
}
\end{figure}

\clearpage

\begin{figure}[p]
\includegraphics{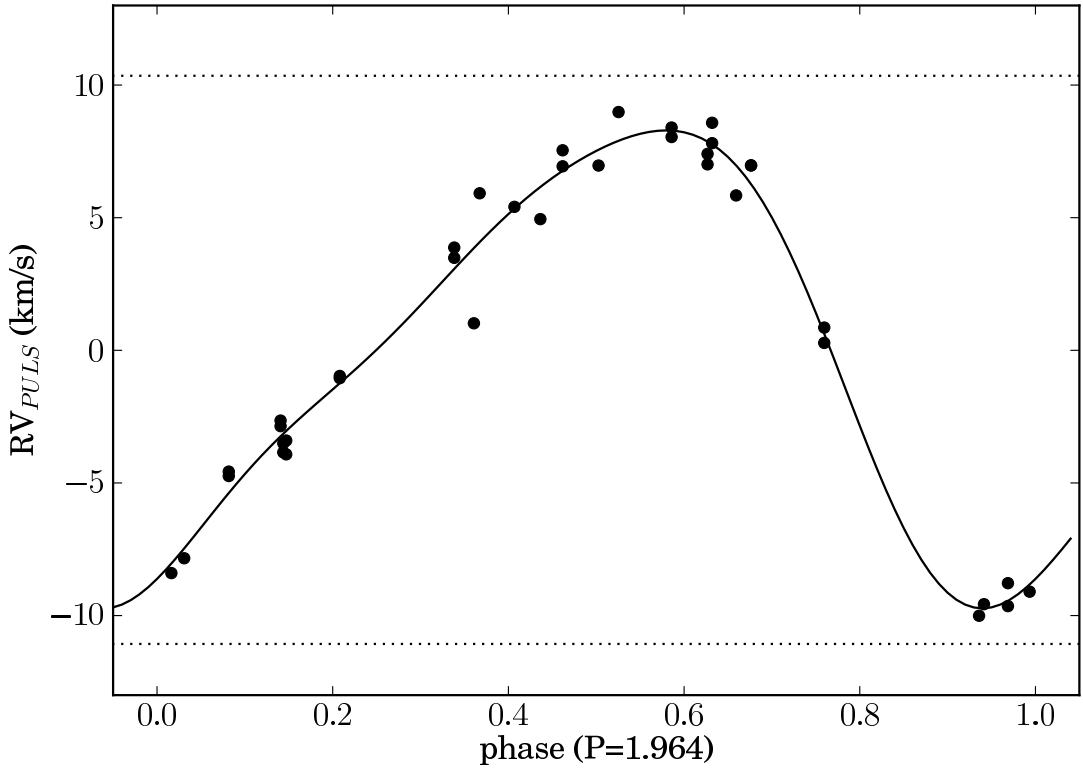}
\vspace{10cm}
\caption{Pulsational radial velocity curve of the primary component of the OGLE-LMC-CEP-1718 system.
The pulsational phases of the Cepheid are well covered by the observations. The solid curve is a
Fourier series fit to the data. The two horizontal dotted
lines indicate the radial velocity amplitude of the secondary, longer-period Cepheid in the system.
}
\end{figure}

\clearpage

\begin{figure}
\includegraphics{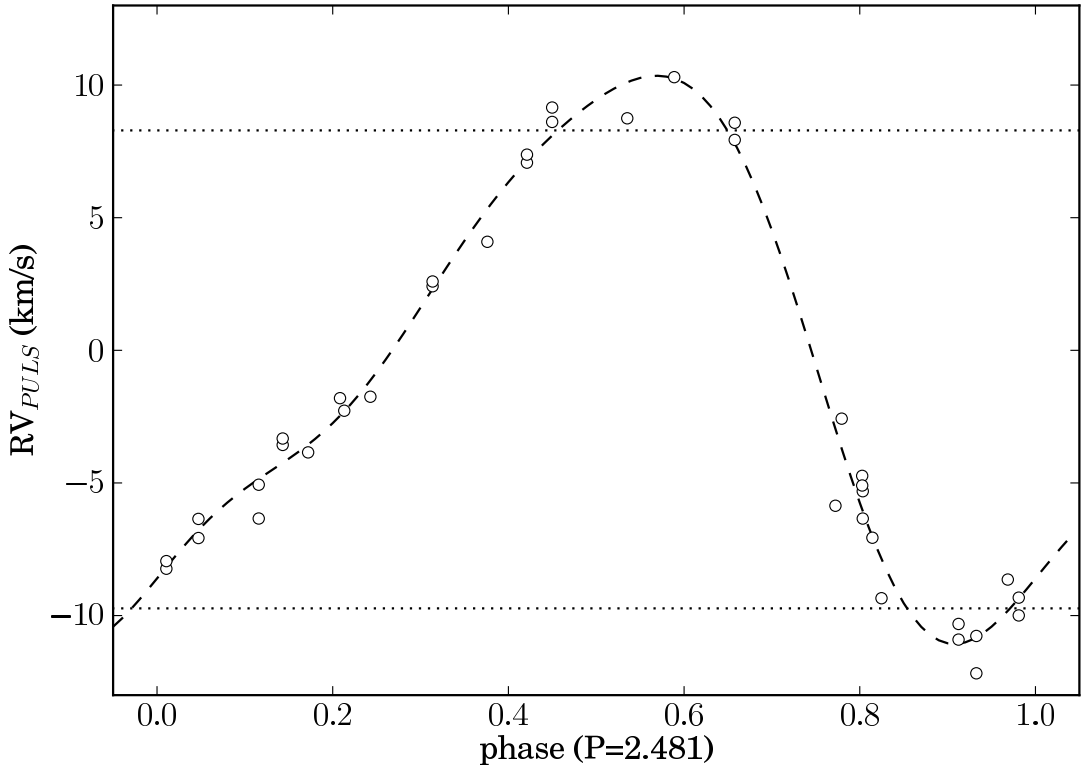}
\vspace{10cm}
\caption{Pulsational radial velocity curve of the secondary component of the OGLE-LMC-CEP-1718 system.
The data cover the pulsation cycle of the Cepheid very well. The dashed line is a Fourier series fit
to the data. The horizontal dotted lines indicate the (smaller) radial velocity amplitude of the
primary Cepheid in the system.
}
\end{figure}

\clearpage

\begin{figure}
\includegraphics{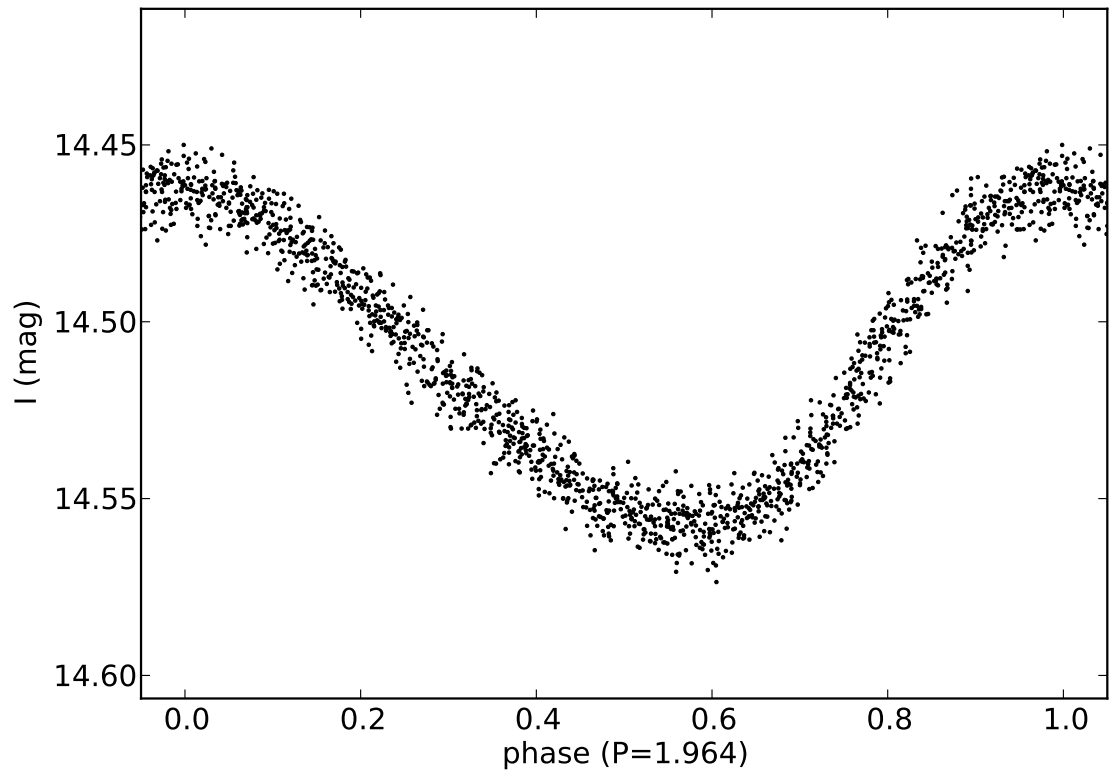}
\vspace{10cm}
\caption{Pulsational I-band light curve of the primary component of the OGLE-LMC-CEP-1718 system.
The small amplitude and the near-sinusoidal shape of the light curve are typical for Cepheids
pulsating in the first overtone mode.
}
\end{figure}

\clearpage

\begin{figure}
\includegraphics{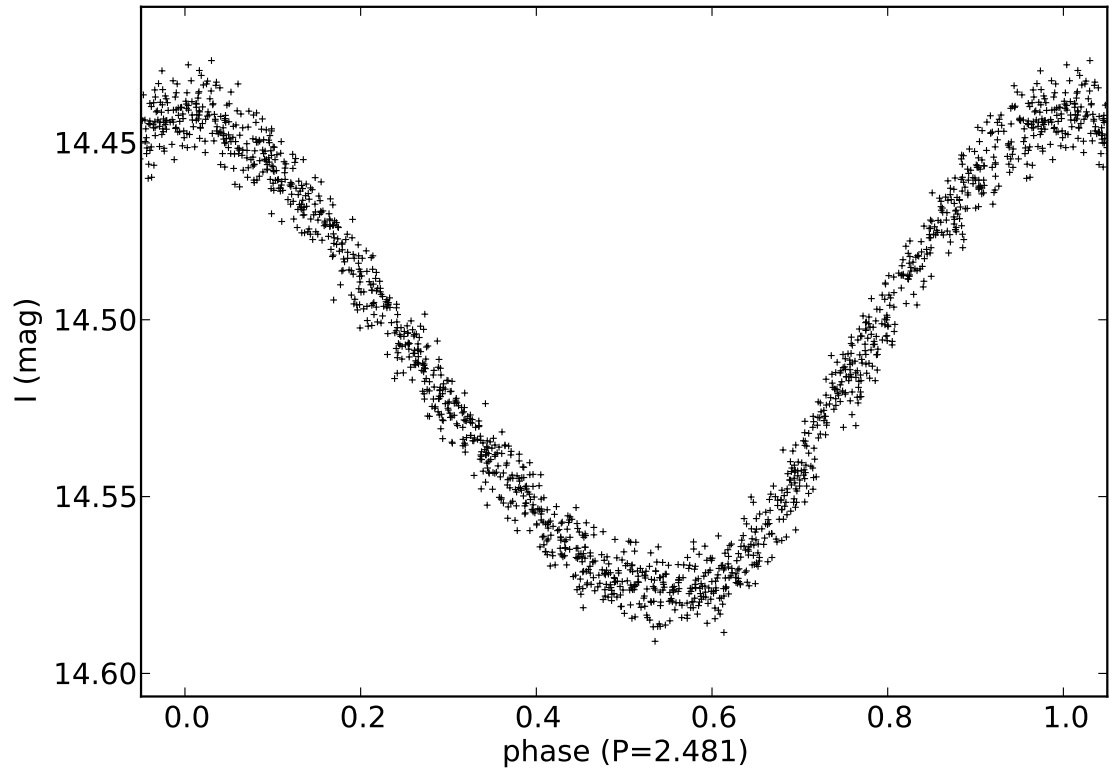}
\vspace{10cm}
\caption{Pulsational I-band light curve of the secondary component of the OGLE-LMC-CEP-1718 system.
As for the primary component, the light curve amplitude and symmetry suggest that the secondary Cepheid
is a first overtone pulsator as well.
}
\end{figure}

\clearpage

\begin{figure}
\includegraphics{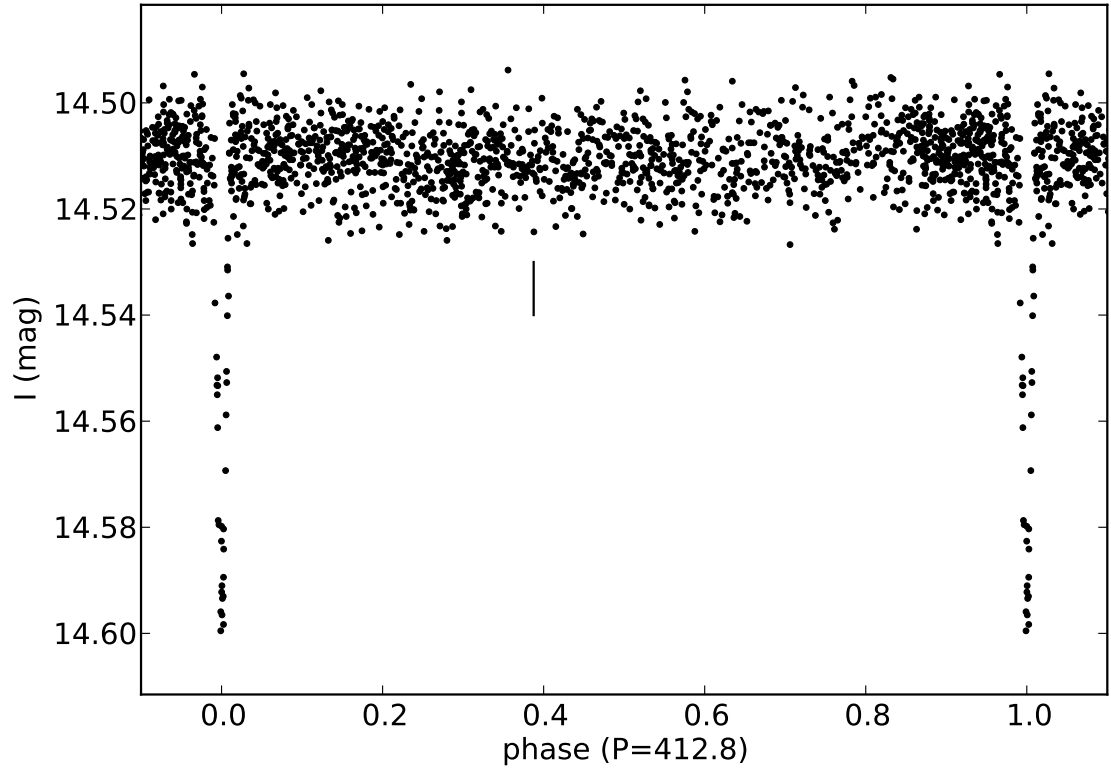}
\vspace{10cm}
\caption{Orbital I-band light curve of the OGLE-LMC-CEP-1718 system, with the pulsational variabilities
of its two Cepheid components removed. Only the primary eclipse is visible. The bar below the light curve indicates
the expected position of the secondary eclipse as estimated from the orbital solution given in Table 2.
}
\end{figure}

\clearpage

\begin{figure}
\includegraphics{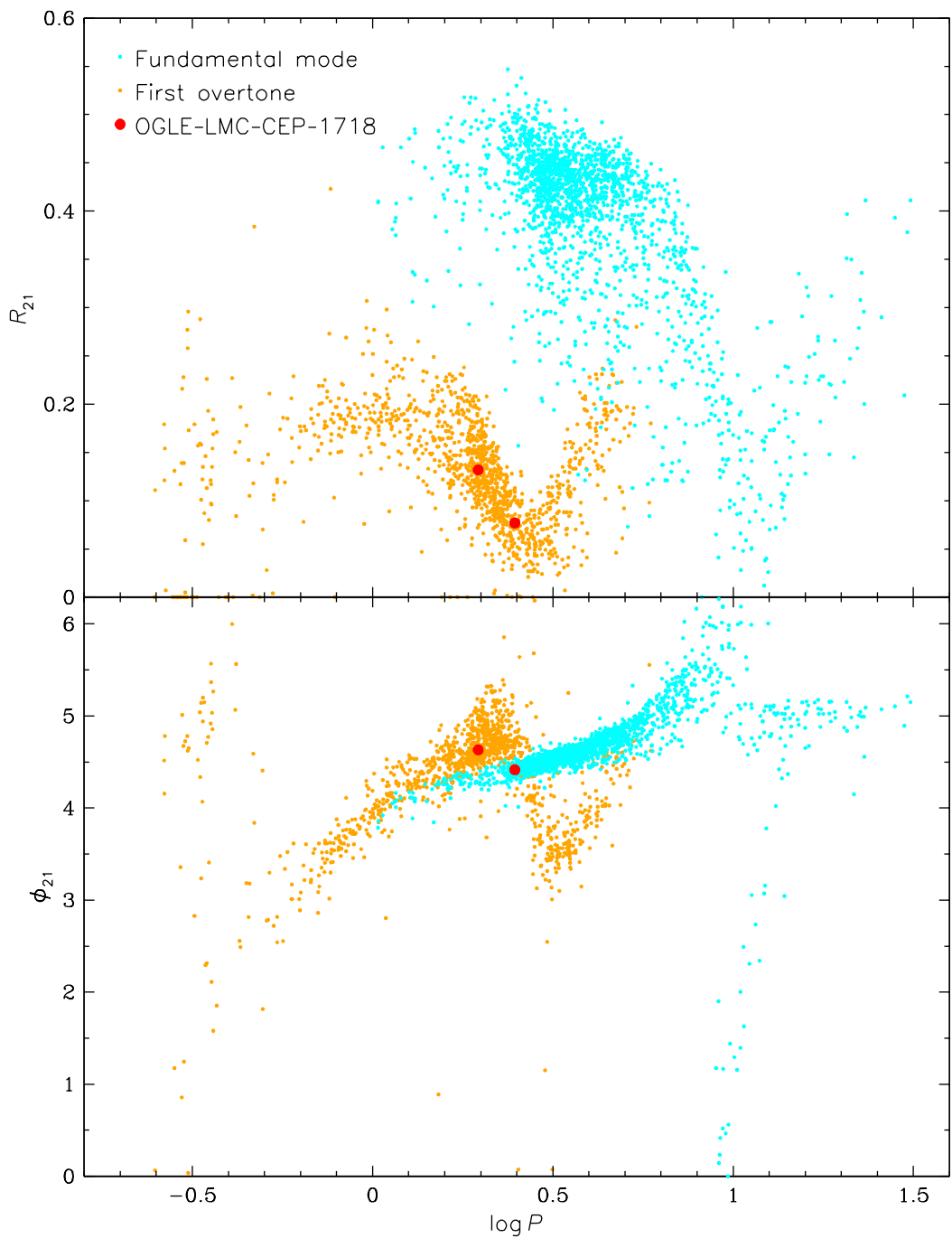}
\vspace{18cm}
\caption{Fourier decomposition parameters from the pulsational I-band light curves of the two
Cepheids in the OGLE-LMC-CEP-1718 system, plotted against the log of the observed pulsation periods (in days).
Both Cepheids  clearly lie on the sequence for first overtone pulsators (yellow points; data from Soszynski
et al. 2008) in the amplitude-period plot (upper panel). In the lower panel phase-period plane, the mode
identification for the secondary (P=2.48 day) Cepheid is complicated because it lies in the overlapping part
of the first overtone and fundamental mode sequences in this diagram. Its position in the upper panel diagram
however allows a very secure identification of its first overtone pulsation mode.
}
\end{figure}

\clearpage

\begin{deluxetable}{ccccccccc}
\tablecaption{Radial velocities of the OGLE-LMC-CEP-1718 system.\label{tab:sup}}    
\tablewidth{0pt}
\tablehead{
\colhead{HJD} & \colhead{$RV_1$} & \colhead{$RV_2$} & \colhead{HJD} & \colhead{$RV_1$} & \colhead{$RV_2$} & 
\colhead{HJD} & \colhead{$RV_1$} & \colhead{$RV_2$}
}
\startdata
  5833.7700 &  227.2 &  256.7 &  5965.7124 &  282.4 &  214.1 &  6573.8505\tablenotemark{a} &  218.8 &  269.7 \\
  5833.7701 &  226.7 &  255.3 &  6309.7243 &  278.5 &  225.7 &  6577.7492\tablenotemark{a} &  217.8 &  264.8 \\
  5833.8898 &  229.7 &  257.4 &  6314.7440 &  268.2 &  220.1 &  6637.5343 &  233.3 &  261.4 \\
  5833.8898 &  229.6 &  258.1 &  6314.7440 &  268.4 &  219.7 &  6637.5343 &  232.7 &  262.0 \\
  5836.7729 &  238.6 &  262.9 &  6529.9196\tablenotemark{a} &  239.4 &  273.2 &  6637.7779 &  234.2 &  264.0 \\
  5836.7729 &  238.6 &  263.1 &  6554.8613\tablenotemark{a} &  230.2 &  279.3 &  6637.7779 &  233.9 &  264.3 \\
  5951.5845 &  278.4 &  222.5 &  6558.8792 &  234.0 &  267.7 &  6637.8685 &  233.6 &  265.2 \\
  5951.5845 &  278.7 &  222.8 &  6560.7652 &  234.3 &  261.1 &  6637.8685 &  234.4 &  265.9 \\
  5952.5329 &  289.1 &  209.3 &  6571.7000\tablenotemark{a} &  216.9 &  275.3 &  6638.5298 &  216.3 &  274.6 \\
  5952.5329 &  289.5 &  210.3 &  6572.6830\tablenotemark{a} &  231.7 &  265.5 &  6638.5298 &  217.2 &  274.8 \\
  5964.5759 &  279.2 &  226.8 &  6572.8131\tablenotemark{a} &  233.7 &  262.0 &  6638.8675 &  223.3 &  280.7 \\
  5964.5759 &  279.8 &  227.4 &  6573.6746\tablenotemark{a} &  217.1 &  267.6 &  6638.8675 &  223.1 &  281.3 \\
  5965.7124 &  282.0 &  212.8 &  6573.7767\tablenotemark{a} &  217.5 &  269.2 &            &        &        \\
\enddata
\tablenotetext{a}{obtained using HARPS data (all the rest -- using MIKE data). The radial velocities are
in km/s.}
\end{deluxetable}

\clearpage

\begin{deluxetable}{lr@{ $\pm$ }l}
\tablecaption{Orbital solution for CEP-1718.} 
\tablewidth{0pt}
\tablehead{
\colhead{Parameter} & \multicolumn{2}{c}{Value}
}
\startdata
$\gamma$ (km/s)       &  248.97  & 0.15 \\ 
$T_0$ (d)         &  2450697.3   & 0.9 \\
$a \sin i$ ($R_\odot$)&  452.8    & 3.5  \\ 
$q=M_2/M_1$           &    0.993  & 0.013 \\ 
$e$                   &    0.276  & 0.013 \\
$\omega$ (deg)        &  308.6    & 1.8  \\
$K_1$ (km/s)          &   28.76   & 0.25 \\ 
$K_2$ (km/s)          &   28.96   & 0.30 \\ 
\enddata
\end{deluxetable}

\clearpage

\begin{deluxetable}{lcc}
\tablecaption{Properties of CEP-1718. \label{tab:abs}}
\tablewidth{0pt}
\tablehead{
\colhead{Parameter} & \colhead{Primary} & \colhead{Secondary}
}
\startdata
pulsation period (d) & 1.9636625 & 2.480917 \\
pulsation mode  & FO & FO \\
mass ($M_\odot$) & 3.3 $\pm$ 0.11 & 3.28 $\pm$ 0.11 \\ 
radius\tablenotemark{a} ($R_\odot$) & 24.0 $\pm$ 1.2  & $28.5_{-1.1}^{+2.9}$ \\ 
orbital period (d) & \multicolumn{2}{c}{412.807 $\pm$ 0.008 } \\
$T_{pri}$ (d)    & \multicolumn{2}{c}{2455050.5  $\pm$ 0.1 } \\
semimajor axis ($R_\odot$) & \multicolumn{2}{c}{454.9 $\pm$ 3.6} \\
inclination (deg) & \multicolumn{2}{c}{$84.5_{-0.4}^{+0.15}$} 
\enddata
\tablenotetext{a}{
The values were calculated using the sum of the radii from our light curve analysis,
 and the ratio obtained from the period - radius relation for first overtone Cepheids
 of Sachkov (2002). The errors were estimated independently.
}
\end{deluxetable}

\end{document}